\documentclass[showpacs,twocolumn,amsmath,amssymb,prb]{revtex4}

\usepackage{graphicx}
\usepackage{amsmath,amsfonts,amssymb,amsthm}
\usepackage{fancyhdr}
\usepackage{mathrsfs}

\usepackage[latin1]{inputenc}
\usepackage[colorlinks]{hyperref}
\begin{document}

\title{Phase-dependent quasiparticle tunneling in Josephson junctions: Measuring the $\cos\varphi$-term with a superconducting charge qubit}
\author{Juha Lepp\"akangas, Michael Marthaler, and Gerd Sch\"on}

\affiliation{Institut f\"ur Theoretische Festk\"orperphysik
and DFG-Center for Functional Nanostructures (CFN), Karlsruhe Institute of Technology, D-76128 Karlsruhe, Germany}


\begin{abstract}
 We investigate quasiparticle tunneling in a Cooper-pair box which is embedded
 in a superconducting ring to allow control of the total phase
 difference across the island. The phase affects the transition rate 
 between different electron number parity states of the island,
 which can be observed in experiment by established means.
 The phase dependence also leads to what is known as the $\cos\varphi$-term
 in the tunneling characteristics of classical Josephson junctions. This effect has remained 
 controversial for decades; the proposed scheme opens an independent way to probe it.
\end{abstract}

\maketitle

{\bf Introduction.}
 Since the discovery of the Josephson effect~\cite{Josephson,Anderson}
 it has been known that, in addition to the phase-dependent Cooper-pair current
 and the well-known quasiparticle current,
 there exists a phase-dependent quasiparticle-pair interference term~\cite{Josephson,Harris}. 
 It is often called the $\cos\varphi$-term since in the simplest approximation 
 it adds a dissipative term proportional to $\cos\varphi$ to the 
  RCSJ model of a current-biased Josephson junction, 
 \begin{equation}
 \label{eq_RSCJ-model_with_cos_term}
  C\ddot{\varphi}+\frac{1}{R}(1+\epsilon\cos\varphi)\dot{\varphi}
  +\frac{2e}{\hbar}I_{\rm J}\sin\varphi=\frac{2e}{\hbar}I \,.
 \end{equation}
 Here $C$ is the junction capacitance, $R$ the subgap resistance,
 $I_{\rm J}$ the critical current, and $I$ the  bias current.
 The theory, based on the BCS and tunneling Hamiltonian predicts for slow variations of $\varphi$
 the value
 $\epsilon\approx 1$~\cite{Harris}, whereas various experiments suggested 
 $\epsilon\approx -1$~\cite{Pedersen,Falco,Vincent,Nisenoff,Soerensen}. 
However, these experiments were difficult to interpret, and it is probably fair to say 
 that the discrepancy and its origin was never fully resolved.
 Recently, after a period of reduced interest,
 the role and properties of quasiparticle tunneling in Josephson junctions has
 again attracted attention, since it provides a possibly important contribution to 
 the relaxation in superconducting qubits\cite{Martinis,MartinisNew,DevoretNew,Lutchyn}.

\begin{figure}[t]
\begin{center}
\includegraphics[width=\linewidth]{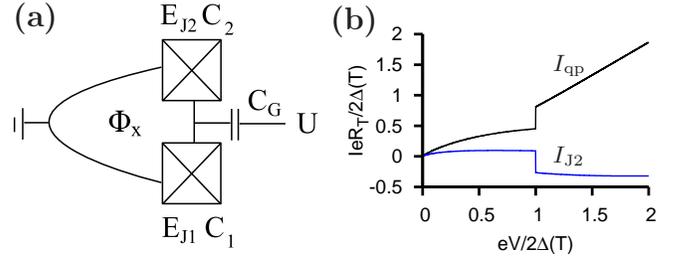}
\caption{(a)
 A Cooper-pair box between two  Josephson junctions in a superonducting loop 
 threaded by a flux $\Phi_x$. The small island is capacitively coupled 
 to a voltage gate $U$.
(b) The quasiparticle current $I_{\rm qp}$ and the quasiparticle-pair
interference term $I_{\rm J2}$
for a voltage-biased Josephson junction with tunneling resistance $R_{\rm T}$ at $T=0.9T_{\rm c}$.
}
\label{fig1}
\end{center}
\end{figure}

 In this article we show that the quasiparticle-pair interference term
 also affects the single-electron tunneling rate in low-capacitance superconducting tunnel junctions. 
 We propose a scheme to measure it via transitions between different charge eigenstates of a
 Cooper-pair box  (CPB) embedded in a superconducting ring as shown in Fig.~\ref{fig1}a.
 The system is particularly sensitive to quasiparticle tunneling, as these processes 
 change the particle-number parity of the island.
 The signature of the quasiparticle-pair interference is a transition rate with
 additional terms proportional to $\epsilon\cos\varphi_i$, where
 the phases $\varphi_i$ can be controlled by the applied magnetic flux,
 allowing tuning the effect~\cite{DevoretNew}.
 States with even or odd parity can be distinguished by measuring the effective quantum 
 capacitance of the CPB, as demonstrated in the experiments of Ref.~\onlinecite{Duty}. 
 This technique has already been used to probe quasiparticle tunneling in the presence of their
 typical non-equilibrium density, but no attention 
 has been paid to the phase dependence and the interference term~\cite{Shaw}.


 One of the difficulties of the experiments aimed at resolving the $\cos \varphi$-term lies in the fact 
 that with current-biased junctions one can control and probe the phase-dependent quality factor 
 of small oscillations only for a restricted range of phase values.
 SQUID configurations~\cite{Vincent,Nisenoff} where
 the flux can be biased and the conductance can be measured
 over the whole phase periodicity offer, in principle, a solution to this problem.
 However, such setups suffer from the problem that variations in $\varphi$ change,
 simultanously with the dissipation due to the $\cos \varphi$-term, also  the oscillation frequency.
 This in turn can alter the effect of the electromagnetic environment,
 which in many cases is the main source of dissipation~\cite{TinkhamBook}.
 Our proposed measurement scheme overcomes these problems as
 for the CPB the dominant energy scale, the charging energy, is independent of the flux,
 and for the change of particle-number parity of the island
 no significant process competing with quasiparticle tunneling exists.

\begin{figure}[t]
\begin{center}
\includegraphics{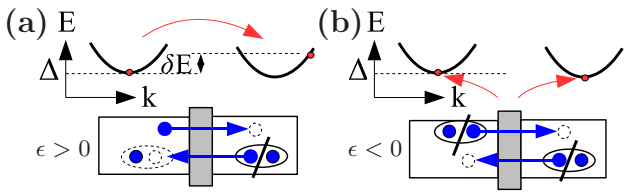}
\caption{(a)
 A quasiparticle tunneling across the Josephson junction from left to right
 gains the even-odd energy difference $\delta E$. It can
 involve either electron tunneling from left to right (upper process), or 
 electron tunneling from right to left (lower). 
 In the latter process a Cooper pair breaks on the right
 and the tunneling electron recombines with an unpaired electron on the left.
 (b)
 Breaking of a Cooper pair during quasiparticle tunneling. 
 This process sets in when $\delta E>2\Delta$ and has also two possible electron tunneling directions.}
\label{fig2}
\end{center}
\end{figure}

 A possible reason for the sign discrepancy of $\epsilon$ lies in the 
 dependence of the quasiparticle-pair interference term on the voltage $V\propto \dot{\varphi}$. 
 For constant voltage it can be easily evaluated, the result $I_{\rm J2}$ is
 plotted in Fig.~\ref{fig1}b. 
 It changes sign at a voltage matching the superconducting gap, $eV=2\Delta$.
 Accordingly, $\epsilon$ is expected to be positive below the gap and and negative above.
 A broadening of the quasiparticle states can lead to
 a negative sign also at subgap voltages~\cite{Zorin,Likharev}.
 However, for typical materials used today this effect should be negligible.
 In the setup of Fig.~\ref{fig1}a, without further modification, 
 only thermally excited or non-equilibrium quasiparticles can tunnel (see Fig.~\ref{fig2}a).
 This is a subgap process corresponding to positive sign for $\epsilon$.
 In order to probe the sign change of $\epsilon$ we propose in the second part of this article
 a modified setup with a lower superconducting energy gap in part of the ring. 
 Parameters are chosen such that enough energy is available to split a Cooper-pair
 (see Fig.~\ref{fig2}b) in a quasiparticle tunneling process across one of the JJs.
 This corresponds to a process above the gap and, hence, to a negative sign for $\epsilon$.



{\bf System.}
The system considered is shown in Fig.~\ref{fig1}a. It
consists of a small island (Cooper-pair box) capacitively coupled to a gate
and embedded via two Josephson junctions in a superconducting ring 
that is threaded by an external magnetic 
flux $\Phi_x=(\hbar/2e)\phi_x$.
If the inductance of the ring $L$ is  small,
we can assume that the
phase difference across the split CPB is equal to the phase bias $\phi_x$.
The coherent time evolution of the system is then described by the
Hamiltonian
\begin{equation}
 H_{\rm coh}=E_C(N-N_{\rm G})^2-E_{\rm J 1}\cos(\varphi+\phi_x)-E_{\rm J 2}\cos(\varphi),
\label{coherenthamiltonian}
\end{equation}
where $E_{{\rm J} i}$ are the Josephson couplings of the two junctions,
$E_C=e^2/2C_{\Sigma}$ denotes the charging energy of the island
with $C_{\Sigma}=C_1+C_2+C_{\rm G}$ being the
sum of the capacitances surrounding the island, and $N_{\rm G}=-C_{\rm G}U/e$. The phase
$\varphi$ and 
the number of excess electron charges on the island $N$
satisfy the periodic commutation relation
$[N/2,e^{\pm i\varphi}]=\pm e^{\pm i\varphi}$. 
Cooper pairs tunnel coherently across the two junctions with a relative phase
shift $\phi_x$, the phase difference across the split CPB. It can be treated as a classical variable. 

In the following we assume that $E_C> E_{\rm J 1},E_{\rm J 2}$, and the gate is biased at $N_{\rm G}=1$.
Therefore we restrict our analysis to the subspace spanned by the
island charge states $\vert 0\rangle$, $\vert 1\rangle $, and $\vert 2\rangle $, with $N=0,1$, or $2$
extra electron charges. Choosing the energy of the state $\vert 1\rangle$ as reference
the coherent part of the Hamiltonian takes the form 
\begin{eqnarray}
 H_{\rm coh}&=&E_C\left(\vert 0\rangle\langle 0\vert+\vert 2\rangle\langle 2\vert\right)\nonumber\\ 
&-&\frac{E_{\rm J}}{2}\left(e^{i\varphi_{\rm eff}}\vert 2\rangle\langle 0\vert+e^{-i\varphi_{\rm eff}}\vert 0\rangle\langle 2\vert\right).
\label{coherenthamiltonian2}
\end{eqnarray}
Here we introduced $E_{\rm J}e^{i\varphi_{\rm eff}}=E_{\rm J 1}e^{i\phi_x}+E_{\rm J 2}$,
with $E_{\rm J}=\sqrt{E_{\rm J 1}^2+E_{\rm J 2}^2+2E_{\rm J 1}E_{\rm J 2}\cos\phi_x} \geq 0$ and
$\tan\varphi_{\rm eff}=\sin\phi_x/\left(\cos\phi_x+E_{\rm J 2}/E_{\rm J 1}\right)$.
If $E_{\rm J 1}>E_{\rm J 2}$  the effective flux $\varphi_{\rm eff}$ spans the whole  
range from $-\pi$ to $\pi$.
The Hamiltonian (3) does not connect states with even and odd numbers
of electrons on the island. It has the eigenstates
\begin{eqnarray}
\vert\uparrow\rangle&=&\frac{1}{\sqrt{2}}\left( \vert 0\rangle -e^{i\varphi_{\rm eff}}\vert 2\rangle  \right)\nonumber  \\
\vert\downarrow\rangle&=&\frac{1}{\sqrt{2}}\left( \vert 0\rangle +e^{i\varphi_{\rm eff}}\vert 2\rangle  \right),\label{eq_Eigenstates_at_the_symmetry_point}  
\end{eqnarray}
in addition to the "odd" state $\vert {\rm o}\rangle=\vert 1\rangle$,
with  eigenenergies $E_{\uparrow}=E_C+E_{\rm J}/2$,
$E_{\downarrow}=E_C-E_{\rm J}/2$ and $E_{\rm o}=0$. 
The relevant energy difference in the following is the one between the two lowest eigenstates
$\delta E=E_{\downarrow}$.

The quasiparticles are described by the BCS  Hamiltonian.
Their tunneling across the junctions leads to transitions between states having even or odd number
of electrons on the island. For junction $i$ the tunneling is described by the Hamiltonian
\begin{eqnarray}\label{eq_Tunneling_Hamiltonian_Definition}
&&H_{{\rm T} i}=t_i\sum_{{\bf kl}}[(u_{\bf{k}}\gamma^{\dagger}_{{{\bf{k}}}\uparrow}+v_{\bf{k}}\gamma_{{{\bf{k}}}\downarrow })(u_{\bf{l}}\gamma_{{{\bf{l}}}\uparrow }+v_{\bf{l}}\gamma_{{{\bf{l}}}\downarrow }^{\dagger})\hat T_i\nonumber\\
&+&(v_{\bf{k}}\gamma^{\dagger}_{{{\bf{k}}}\uparrow}-u_{\bf{k}}\gamma_{{{\bf{k}}}\downarrow })(u_{\bf{l}}\gamma_{{{\bf{l}}}\downarrow }^{\dagger}-v_{\bf{l}}\gamma_{{{\bf{l}}}\uparrow})\hat T_i^{\dagger}+{\rm h.c.}],
\end{eqnarray}
where $\gamma_{{\bf l}\sigma}^{(\dagger)}$  is a quasiparticle annihilation (creation) operator of the state $\bf l$ and spin $\sigma$ in the island.
The states in the loop side of the junctions are labelled as $\bf k$.
The quasiparticle operators are related to the corresponding electron operators via the Bogoliubov transformation
$c_{{\bf{l}}\uparrow}^{\dagger}=u_{\bf{l}}\gamma^{\dagger}_{{{\bf{l}}}\uparrow}+v_{\bf{l}}\gamma_{{{\bf{l}}}\downarrow}$ and
$c_{-{\bf{l}}\downarrow}^{\dagger}=-v_{\bf{l}}\gamma_{{{\bf{l}}}\uparrow}+u_{\bf{l}}\gamma^{\dagger}_{{{\bf{l}}}\downarrow}$.
Here the BCS coherence factors $u$ and $v$
are real numbers as the phase dependence is included
with the charge-transfer operators $\hat T_2=\vert 2 \rangle\langle 1\vert+\vert 1\rangle\langle 0 \vert$ and $\hat T_1=e^{i\phi_x/2}\hat T_2$, 
which take care of the corresponding island-charge changes in the tunneling events. They have
similar (but halved) external flux dependence as the Cooper-pair tunneling, described by Eqs.~(\ref{coherenthamiltonian}-\ref{coherenthamiltonian2}).
The average tunneling matrix elements $t_i$
are related to the normal state tunnel resistances
via $R_{{\rm T} i}=\hbar/ t_i^2  D^2\pi e^2$, 
where $D$ is the density of states at the Fermi surface, including spin.


{\bf Tunneling processes.}
The tunneling Hamiltonian (\ref{eq_Tunneling_Hamiltonian_Definition}),
when treated perturbatively in leading order,
includes three distinct types of quasiparticle tunneling processes. In the first  
a quasiparticle, either thermally excited or a non-equilibrium one,
tunnels across the junction.
The process can still involve two electron-tunneling directions (terms like $\gamma_{{\bf k}}^{\dagger}\gamma_{{\bf l}}T$ or
$\gamma_{{\bf k}}^{\dagger}\gamma_{{\bf l}}T^{\dagger}$), as visualized in Fig.~\ref{fig2}a. The second process
needs at least an energy $2\Delta$ to break a Cooper pair, which then provides an electron that
tunnels across the barrier (terms like $\gamma_{{\bf k}}^{\dagger}\gamma_{{\bf l}}^{\dagger}T$
and $\gamma_{{\bf k}}^{\dagger}\gamma_{{\bf l}}^{\dagger}T^{\dagger}$ illustrated in Fig.~\ref{fig2}b). 
For the moment we can ignore it, but it plays a role in the modified design  discussed later.
The third tunneling type corresponds to the case when an electron tunnels across a junction,
with a simultaneous recombination of two excitations originating from different
sides of the junction (terms like $\gamma_{{\bf k}}\gamma_{{\bf l}}T$ and $\gamma_{{\bf k}}\gamma_{{\bf l}}T^{\dagger}$). The process does not contribute under the operating conditions considered here.

The even eigenstates are superpositions of
the two charge states $N=0$ and $N=2$, as given by Eq.~(\ref{eq_Eigenstates_at_the_symmetry_point}).
Therefore, in the transition
from, e.g., $\vert\downarrow\rangle$ to $\vert {\rm o}\rangle$
both electron tunneling directions contribute.
They are coherent, and hence the two amplitudes have to be added before squaring in the Golden rule calculation. This introduces an extra contribution to the transition rate, which
is proportional to the cosine of the relative phase. It  corresponds to the $\cos\varphi$-term in single Josephson junctions mentioned in the introduction.
This contribution would not be contained in the standard semiconductor model
treatment~\cite{TinkhamBook}.
Moreover, the interference term depends on which tunneling process is dominant. 
It differs for the tunneling of an existing quasiparticle from the process which involves
breaking of a Cooper-pair, the difference being in the sign of $\epsilon$ of the interference term.


{\bf Transition rates.}
In the following we assume $E_{\rm J}\gg k_{\rm B}T$, which implies that of the two even 
 parity states
only the lower eigenstate $\vert\downarrow\rangle$ needs to be considered.
The quantity to be calculated is the transition rate $\Gamma_{\rm T}$ from this state 
to the odd ground state $\vert \rm o\rangle$. 
The  opposite transition from odd to even occurs either through tunneling
of thermally excited quasiparticles or of a non-thermalized quasiparticle, as discussed in Ref.~\onlinecite{Lutchyn}.
These two odd-to-even processes also show 
interference, but the rate depends strongly on the energy barrier $\delta E$, which
can have a further dependence on the magnetic flux through $E_{\rm J}$.
In order to avoid this complication and to observe only the interference effect we 
concentrate on the even-to-odd transition. It can be resolved in the experiment where
the transitions are seen in real time as jumps
in the phase of a reflected weak microwave signal applied to the gate, due to simultaneous
changes in the quantum capacitance of the system~\cite{Shaw}.
One of the first quantum capacitance observations has been achieved using
the exact same setup we consider in this paper~\cite{Sillanpaa}.


The total even-to-odd transition rate can be decomposed into four contributions,
\begin{equation}
\Gamma_{\rm T}=\Gamma_{\rm 1}^+ +\Gamma_{\rm 1}^- +\Gamma_{\rm 2}^+ +\Gamma_{\rm 2}^-,
 \label{TotalRate}
\end{equation}
where $\Gamma_{1(2)}^{\pm}$ is the forward/backward rate for quasiparticle tunneling through 
junction 1 and 2, respectively.
Expanding the time evolution of the density matrix
up to second order in the tunneling Hamiltonian (\ref{eq_Tunneling_Hamiltonian_Definition})
we obtain the forward tunneling rate for junction 1 
\begin{eqnarray}
\Gamma_{1}^+&=&\frac{1}{R_{{\rm T}1}e^2}\int_{-\infty}^{\infty}\int_{-\infty}^{\infty}d\xi_k d\xi_l \, P_{1}(\xi_k,\xi_l)\,\nonumber \\ 
&\times& f_{\rm L}(\xi_k)[1-f_{\rm I}(\xi_l)]\delta(E_k-E_l+\delta E).
 \label{CPBRates}
\end{eqnarray}
Here $E_{k(l)}=\sqrt{\Delta^2+\xi_{k(l)}^2}$ is the quasiparticle excitation energy and $f_{\rm L(I)}[\xi_{k(l)}]$ the excitation occupation
probability in the loop (island) side of junction 1.
Similar expressions apply for the backward tunneling rate and junction 2.
The information of the interference is contained in the function
\begin{gather}
P_{1(2)}=\frac{1}{4}\left[ 1-\cos \varphi_{1(2)} \frac{\Delta^2}{E_kE_l} +\frac{\xi_k\xi_l}{E_kE_l} \right]\label{interference},
\end{gather}
where $\varphi_1=\phi_x-\varphi_{\rm eff}$ and $\varphi_2=\varphi_{\rm eff}$.
The first term corresponds to the quasiparticle current, while the $\cos\varphi$-term is
the quasiparticle-pair interference~\cite{Harris}.
The last term gives no contribution if the distribution is symmetric
with respect to electron ($k>k_{\rm F}$) and hole ($k<k_{\rm F}$) branches, which is the case for thermal distribution.

To proceed we use a quasiparticle distribution
$f_{\delta\mu}(E) =f_{\rm eq}(E-\delta\mu)$
where $\delta\mu$ is a shift in the chemical potential and $f_{\rm eq}(E)$ the equilibrium Fermi function. 
This description in terms of a shifted chemical potential implies that
all quasiparticles are in thermal equilibrium,
but their density is out of equilibirum ($\delta\mu> 0$).
This type of non-equilibrium distribution may emerge, e.g., in the presence of
an electromagnetic radiation~\cite{OwenScalapino}, and it has been realized in recent experiments~\cite{Palmer,Shaw}.
It has also been used to model quasiparticles in the odd parity state~\cite{Zaikin}.
We further assume that the quasiparticles are distributed uniformly 
in the system and, hence, $\Gamma_{1(2)}^-=\Gamma_{1(2)}^+$.
At low temperatures the decay rate~(\ref{TotalRate}) then reduces to
\begin{gather}
\Gamma_{\rm T}=\sum_{i=1}^2\frac{n_{\rm qp}}{DR_{{\rm T}i}e^2}\left(1-\epsilon\cos\varphi_i\right)\frac{1+r}{\sqrt{r^2+2r}},
\label{approximation1}
\end{gather}
where $r=\delta E/\Delta$, $\epsilon=1/(1+r)$, and $n_{\rm qp}=2D\int_{\Delta}^{\infty}dEf_{\delta\mu}E/\sqrt{E^2-\Delta^2}$
is the quasiparticle density.

\begin{figure}[t]
  \begin{center}
  \includegraphics{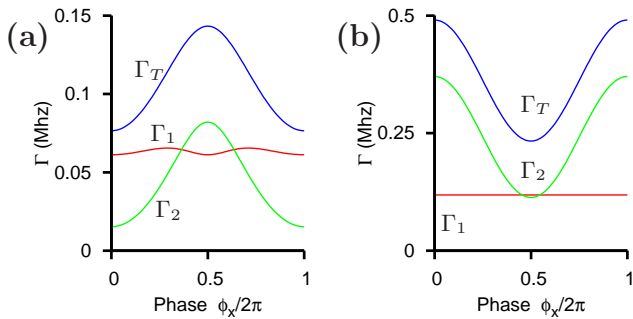}
 \end{center}
\caption{(a)
The transition rate $\Gamma_{\rm T}$ for the even-to-odd particle number switching
as a function of the phase bias $\phi_x$.
Also plotted are the individual contributions for junction 1 ($\Gamma_1=\Gamma_1^++\Gamma^-_1$)
and junction 2 ($\Gamma_2$).
At the considered low temperature, $T=50$~mK,
the transition is dominated by non-equilibrium quasiparticles, for
which we assume the density $n_{\rm qp}=10/\mu {\rm m}^3$, similar to recent experiments\cite{Shaw}.
We use typical aluminum density of states $D\Delta=2.8\times 10^6/\mu{\rm m}^3$, and assume the
Ambegaokar-Baratoff relation between $E_{{\rm J}i}$ and $R_{{\rm T}i}$.
Other parameters are $E_{\rm J1}=4E_{\rm J2}=\Delta/10=E_C/5=20$~$\mu$eV.
(b)
The rates for the modified system with a reduced energy gap in parts of the loop.
The tunneling rate in junction 1
does not depend on the applied flux and the rate of junction 2 has a
sinusoidal variations, but with opposite oscillation phase and hence sign of $\epsilon$
as compared to (a).
The parameters are the same as in (a) except for
$E_{\rm J2}=5*10^{-5}$~$\mu$eV, $E_C=150$~$\mu$eV, and $\Delta_1=4\Delta_2=\Delta$.
}
\label{fig3}
\end{figure}

The transition rate $\Gamma_{\rm T}$ as function of the applied phase $\phi_x$ for typical experimental conditions is plotted in Fig.~\ref{fig3}a.
We have chosen an asymmetric situation with ${\rm min}\{E_{\rm J}(\phi_x)\}/k_B=(E_{\rm J 1}-E_{\rm J 2})/k_{\rm B}\approx 0.2$~K, and the condition $E_{\rm J}\gg k_{\rm B}T$
is satisfied for the chosen temperature $T=50$~mK. We observe an
almost sinusoidal variation of the total rate, which arises from the interference effect.
From the individual contributions of the junctions
we see that this pattern originates from tunneling across the weaker junction 2.
This is the case because for $E_{\rm J 1}\gg E_{\rm J 2}$ one has $\varphi_{\rm eff}\approx\phi_x$, 
and for
quasiparticle tunneling across the junction 1 the first two terms in Eq.~(\ref{interference}) always show destructive interference. For junction 2 the interference
is proportional to $\sim 1-\epsilon\cos\phi_x$, resulting in the sinusoidal phase dependence.
We emphasize that the phase dependence of the total rate originates predominantly 
from the $\cos\varphi$-terms (quasiparticle-pair interference).
A competing effect could be the weak flux dependence of the energy level $\delta E=E_C-E_{\rm J}/2$ through $E_{\rm J}$. However, this contribution remains always small because of the high charging energy of the CPB. The absence of the interference would correspond to $\epsilon=0$.

The temperature dependence of the rates is governed by the quasiparticle density $n_{\rm qp}$.
At high temperatures it is proportional to  $\exp(-\Delta/k_BT)$, at low temperatures it
saturates due the assumed shifted chemical potential $\delta\mu>0$.
Well below this cross-over the {\em individual} junction rates grow (weakly) with decreasing temperature for a constructive interference
and decrease for a destructive one.
However, 
for a system with clearly asymmetric junctions the {\em total} rate always increases with temperature,
since for the larger of the junctions the interference stays practically destructive for all $\phi_x$.


{\bf Sign change of the interference term.}
We now extend our discussion to situations where the  change of sign in $\epsilon$
 can be probed.
For this purpose we assume that the loop consists of two different superconducting materials,
one having a much lower  energy gap than the other.
The structure is chosen such that the gap
changes from $\Delta_1$ to the lower value $\Delta_2\ll \Delta_1$ when crossing junction 1, 
but it remains constant at  $\Delta_2$ 
 across junction 2. The gap has then also to change from $\Delta_2$ back to $\Delta_1$ in the
loop part of the system, for example, via an extra large Josephson junction, which is irrelevant for the parity transitions in the small island.
The interference part (\ref{interference}) is now generalized to
$P_{1(2)}=(1/4)\left[ 1-\cos \varphi_{1(2)} \Delta_{1(2)}\Delta_2/E_kE_l+\xi_k\xi_l/E_kE_l \right]$.
For parameters such that $\Delta_1+\Delta_2>\delta E>2\Delta_2$ Cooper-pairs can be broken in a tunneling process across junction 2.
This leads to an extra term for the junction 2 tunneling of the form
\begin{eqnarray}
\tilde\Gamma_{2}&=&\frac{1}{R_{{\rm T}2}e^2}\int_{-\infty}^{\infty}\int_{-\infty}^{\infty}d\xi_k d\xi_l \, \tilde P_{2}(\xi_k,\xi_l)\, \nonumber\\ 
&\times& [1-f_{\rm L}(\xi_k)][1-f_{\rm I}(\xi_l)]\delta(E_k+E_l-\delta E),
 \label{CPBRates2}
\end{eqnarray}
where $\tilde P_2=1/2-P_2$. Assuming that tunneling across junction 2 dominates we obtain
\begin{gather}
\Gamma_{\rm T}\approx \tilde\Gamma_{2}\approx \frac{\delta E}{R_{\rm T 2}e^2}\left(1-\epsilon\cos\phi_x\right),
\label{approximation2}
\end{gather}
where 
$\epsilon$  equals -1 at
the threshold $\delta E=2\Delta$ and decreases weakly for larger $\delta E$.
In this setup it is essential that $R_{\rm T 2}$ is large enough such that the switching rate does not exceed the qubit splitting $E_{\rm J}$.
For this to be true we need very asymmetric circuit $E_{\rm J 1}\gg E_{\rm J 2}$.
Under these conditions also $\delta E$ has no magnetic flux dependence, and 
 the modulation in the transition rate originates completely from  
 tunneling across junction 2.
Numerical results for the rates are shown in Fig.~\ref{fig3}b.


{\bf Conclusion.}
We have re-analyzed in a new setup the phase dependence of quasiparticle tunneling across 
a Josephson junction.
A split Cooper-pair box embedded in a superconducting ring provides an optimal system to study the effect, as
only quasiparticle tunneling can create transitions between even and odd island charge parity,
and the flux dependence of this rate occurs practically due to the quasiparticle-pair interference.
The transitions can be seen as jumps
in the phase of a reflected weak microwave signal applied to the gate.
The possible presence non-equilibrium quasiparticles makes
the effect observable at very low temperatures, although (especially with differing gaps $\Delta_1\gg\Delta_2$)
it is always possible to increase the thermal tunneling rate by heating the system.
Transitions due to thermal and non-equilibrium quasiparticles result
in a positive sign for $\epsilon$, whereas processes provided by breaking of a Cooper-pair
in a negative one. 
A possible broadening of quasiparticle states would also lead to a negative sign for $\epsilon$.
Therefore, the detection of the discussed phase dependence could also confirm
non-equilibrium quasiparticles as the main source of quasiparticle poisoning in superconducting electronics.


\end{document}